\documentclass[preprint,epsfx,floatfix]{revtex4}
\usepackage{graphicx}
\usepackage{animate}
\usepackage{morefloats}
\usepackage{float}
\usepackage{amsmath}
\begin{document}
\title{
%Heterogeneity facilitates Persistent Infection
Emergence of Persistent Infection due to Heterogeneity
}
\author{Vidit Agrawal$^1$, Promit Moitra$^2$ and Sudeshna Sinha$^2$}
\affiliation{$^1$Department of Physics, University of Arkansas, Fayetteville, Arkansas AR 72701, USA\\
$^2$Indian Institute of Science Education and Research (IISER) Mohali, Knowledge City, SAS Nagar,
  Sector 81, Manauli PO 140 306, Punjab, India}

\begin{abstract}
  We explore the emergence of persistent infection in a patch of
  population, where the disease progression of the individuals is
  given by the SIRS model and an individual becomes infected on
  contact with another infected individual. We investigate the
  persistence of contagion qualitatively and quantitatively, under
  varying degrees of heterogeneity in the initial population. We
  observe that when the initial population is uniform, consisting of
  individuals at the same stage of disease progression, infection
  arising from a contagious seed does not persist. However when the
  initial population consists of randomly distributed refractory and
  susceptible individuals, a single source of infection can lead to
  sustained infection in the population, as heterogeneity facilitates
  the de-synchronization of the phases in the disease cycle of the
  individuals.  We also show how the average size of the window of
  persistence of infection depends on the degree of heterogeneity in
  the initial composition of the population.  In particular, we show
  that the infection eventually dies out when the entire initial
  population is susceptible, while even a few susceptibles among an
  heterogeneous refractory population gives rise to a large persistent
  infected set.

\end{abstract}
\maketitle
\section{Introduction}

How a disease spreads in a population is a question of much interest
and relevance, and consequently has been extensively explored over the
years \cite{McEvedy,kaplan,cliff}.  Mathematically, epidemiological
models have successfully captured the dynamics of infectious disease
\cite{murray,edelstein,CA1,sw,scalefree,network}. One well known model for
non-fatal communicable disease progression is the SIRS cycle. This
model appropriately describes the progression of diseases such as
small pox, tetanus, influenza, typhoid fever, cholera and tuberculosis
\cite{hethcote,ozcaglar}.

The SIRS cycle is described by the following stages. At the outset an
individual is \emph{susceptible} to infection (a stage denoted by S).
On being infected by contact with other infected people in the
neighbourhood, the individual moves on to the \emph{infectious} stage
(denoted by I). This is followed by a \emph{refractory} stage (denoted
by R). In the refractory stage the individual becomes immune to
disease and also does not infect others. But this immunity is
temporary as the individual becomes susceptible again after some time
interval.

Specifically, in this work we consider a cellular automata model of
the SIRS cycle described above \cite{kuperman,gade,kohar}.  In this
model of disease progression, time $t$ evolves in discrete steps, with
each individual, indexed by $(i,j)$ on a $2$ dimensional lattice,
characterized by a counter
\begin{center}
$\tau_{i,j} (t) = 0,1, \dots , \tau_I + \tau_R$
\end{center}
describing its {\em phase} in the cycle of the disease \cite{kuperman}. Here
$\tau_I + \tau_R = \tau_0$, where $\tau_0$ signifies the total length
of the disease cycle. At any instant of time \emph{t}, if phase
$\tau_{i,j}$(t) = $0$, then the individual at site $(i,j)$ is
susceptible; if $1 \le \tau_{i,j}(t) \le \tau_I$, then it is infected;
if phase $\tau_{i,j} (t) > \tau_I$, it is in the refractory stage. For,
phase $\tau_{i,j} (t) \ne 0$ the dynamics is given by the counter updating
by $1$ every time step, and at the end of the refractory period the
individual becomes susceptible again, which implies if $\tau_{i,j} (t)
= \tau_0$ then, $\tau_{i,j} (t+1) = 0$. Namely:
\[
\tau_{i,j} (t+1) =
\begin{cases}
\tau_{i,j} (t) + 1 & \text{ : } 1 \leq \tau_{i,j} (t) < \tau_0\\
0 & \text { : } \tau_{i,j} (t) = \tau_0
\end{cases}
\]

\begin{figure}[ht]
\includegraphics[width=0.7\textwidth]{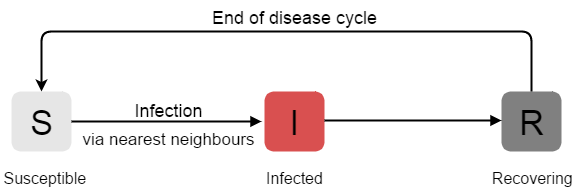}
\caption{Schematic Representation of the SIRS cycle. The color scheme
  in all figures is as follows: black represents the refractory state
  (R); white represents the susceptible state (S); red represents the
  infected state (I).}
\label{schematic}
\end{figure}

Hence the disease progression is a {\em cycle} (see
Fig.\ref{schematic}). We consider the typical condition where the
refractory period is longer than the infective stage,
i.e. $\tau_R > \tau_I$.

%In another direction, Rhodes and Anderson presented a lattice-based
%epidemic model of a non-fatal communicable disease in a mobile host
%population in which they observed novel dynamical behaviour..

We now investigate the spread of epidemic in a group of spatially
distributed individuals, where at the individual level the disease
progresses in accordance with the SIRS cycle. In particular, we
consider a population of individuals on a $2$-dimensional lattice
where every node, representing the individual, has $4$ neighbors
\cite{rhodes}. Unlike many studies with periodic boundary conditions,
here the boundaries are fixed and there are no individuals outside the
boundaries. So our model mimics a patch of population, and
investigates the persistence of infection in such a patch.

%where the individuals are subject to the SIRS disease cycle.

{\em Condition for infection:} Here we consider the condition that a
susceptible individual (S) will become infected (I) {\em if one or
  more of its nearest neighbours are infected}. That is, if $\tau_{i,j}
(t) = 0$, (namely, the individual is susceptible), then
$\tau_{i,j} (t+1) = 1$, if any $1 \le \tau_{x,y} (t) \le \tau_I$
where $x,y \in \{(i-1,j);(i+1,j);(i,j-1);(i,j+1)\}$.

%[Addition]\\
%Coupling equation:
%\begin{align*}
%& \tau_{i,j} (t+1) = 1\ if\  \tau_{i,j} (t) = 0\ and\ 1 \leq \tau_{x,y} (t) \leq \tau_I\\
%& where\ x,y \in \{ (i-1,j), (i+1,j), (i,j-1), (i,j+1)\}
%\end{align*}

\section{Spatio-temporal patterns of infection spreading}

We first focus on the infection spreading patterns in the
population. The principal question we ask is the following: when is
infection persistent in a patch, and how this depends on the
constitution of the initial population. In order to examine this, we
study the spread of infection from a seed of infection (namely one or
two infected individuals) across a patch of population composed of
individuals at different stages in the disease cycle, and with varying
degrees of heterogeneity in the population.

With no loss of generality we consider $\tau_I = 4$; $\tau_R = 9$;
$\tau_0 = 13$ and a lattice of size $100 \times 100$.  In our figures
we represent the state of an individual in the disease cycle (namely S, I
or R) by a color, with white denoting a susceptible individual, black
denoting a refractory individual and red denoting an infected
individual.  The fraction of susceptible individuals in the population
at time $t$ is denoted by $S_t$, the fraction of infected individuals
by $I_t$ and the fraction of refractory individuals by $R_t$. In the
sections below we will focus on the possibility of the prolonged
existence of infection arising in different classes of initial
populations, characterized by different $S_0$, $R_0$ and $I_0$.

% It is obvious that an infected individual surrounded by a completely
% refractory population will not lead to any spread in infection, as
% the infected individual will have no susceptible neighbours while it
% is in the infectious stage, since $\tau_R > \tau_I$. However, it is
% not evident apriori what would happen if the population is
%completely susceptible, or a mixture of susceptible and refractory
%individuals. We investigate these two scenarios below.

%The state of the population at different points in time is
%shown in Figs. 2-8, and it offers a clear insight into how infection
%spreads in the population.

\subsection{Non-persistent Infection in a Homogeneous Susceptible
  Population}

\begin{figure}[ht]
\includegraphics[width=1\textwidth]{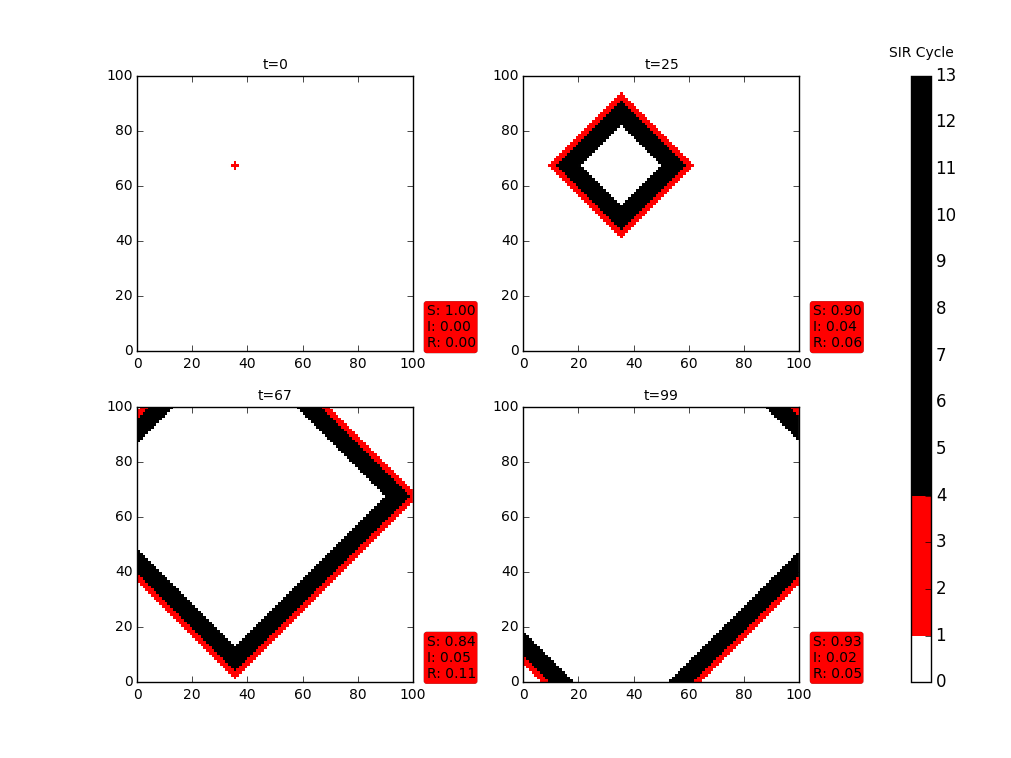}
\caption{Snapshots at times $t= 0, 25, 67, 99$, showing the spread of
  infection from one infected individual at $t=0$, in a homogeneous
  initial population comprising entirely of susceptible individuals
  (i.e. $S_0 \sim 1$, $R_0 = 0$, $I_0 \sim 0$). The long
  bar shows the relative lengths of the susceptible (S), infected (I)
  and refractory (R) stages in the disease cycle, where $\tau_I=4$,
  $\tau_R=9$ and the total disease cycle $\tau_0$ is $13$ (see
  text). The red box shows the fraction of S, I and R individuals in
  the population at that instant of time.}
\label{control}
\end{figure}

First we investigate the effect of an infected individual on a
population patch where {\em all individuals are entirely susceptible
  to infection}. Namely, we consider the case where at the outset
there is one infected individual and the rest of the population is in
the susceptible state, with $\tau_{ij} = 0$. Fig.~\ref{control}
displays the spreading patterns obtained in such a scenario. It is
clearly evident that as time progresses the infection starts from the
infected individual (``seed'') and spreads symmetrically. This
symmetric spreading pattern is readily understood from the condition
for infection, which turns susceptible individuals to infected if any
one of their neighbors is infected. So the infected seed infects its
four neighbors, and these newly infected individuals in turn infect
their nearest four neighbours, and so on. This process leads to an
isotropic wave of infection which stops at the boundaries. We
confirmed the generality of these observations for different relative
lengths of the infectious and refractory periods, namely varying
$\tau_I$ and $\tau_R$ (with $\tau_I < \tau_R$). We further ascertained
that the choice of the location of the infected individual did not
affect these qualitative trends.

Now the key factor in infection spreading is the contact of
susceptible individuals with infected ones. It is clear that such an
interaction takes place only at the outer edge of the wave of
infection, while the inner boundary of the infected zone is contiguous
only to refractory individuals. So the infection only spreads
outwards, and does not move back into the interior of the lattice
again.  Importantly then, the infection is removed after a while from
the patch of population, and all the individuals (including the original
infective seed) comes to the end of the disease cycle and becomes
susceptible again. So there is no infective site left in the
population to perpetuate the infection and initiate another wave of
disease spreading. Thus a {\em fully susceptible population does not
  allow the infection to persist.}

\subsection{Persistent infection in Heterogeneous Populations}

Next we investigate the infection spread in the more realistic
scenario where both refractory ($\tau_{i,j} > \tau_I$) and susceptible
individuals ($\tau_{i,j} = 0$) are present in the initial population,
and are randomly distributed spatially. We first consider the case
where the refractory individuals have phases $\tau_{i,j} = \tau_I
+ 1$, namely, they are at the start of the refractory stage of the
disease cycle. We investigate the persistence of infection in
heterogeneous populations, with the initial state having (a) a single
seed of infection and (b) varying initial fractions of infected
individuals $(I_0)$. In both scenarios, we analyze the effect of
varying $S_0$ and $R_0$ on the persistence of infection.

\begin{figure}[ht]
\includegraphics[width=1\textwidth]{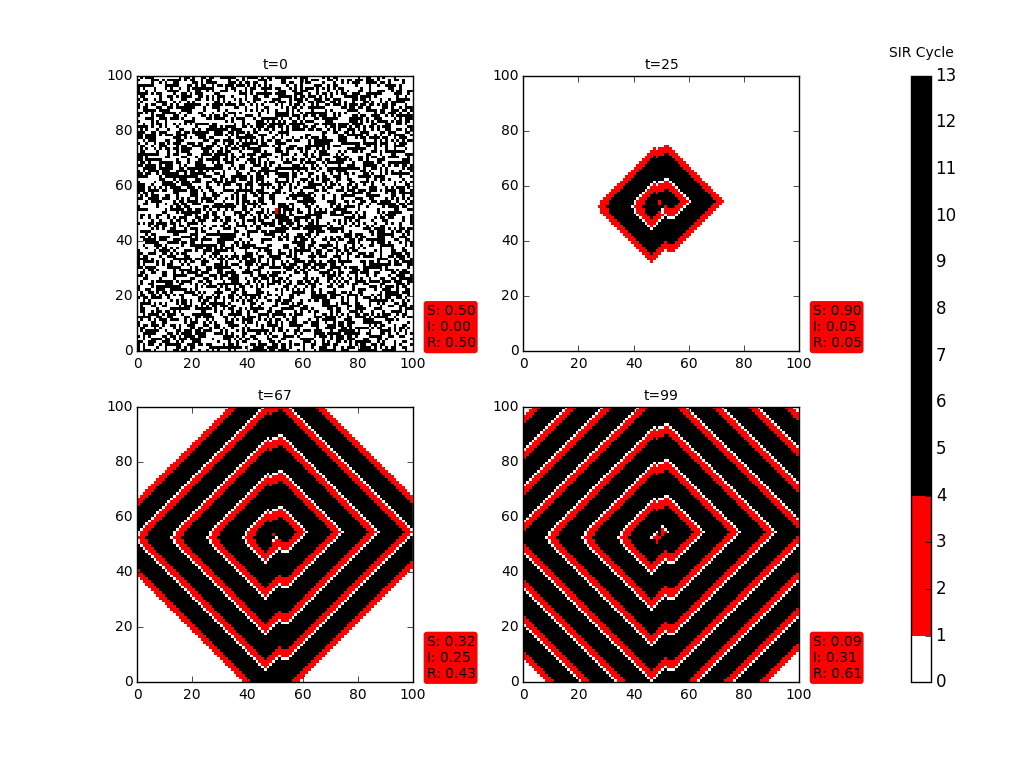}
\caption{Snapshots of the infection spreading pattern in a
  heterogeneous population comprising initially of a random mixture of
  equal numbers of susceptible and refractory individuals ($S_0 \sim
  0.5$, $R_0 \sim 0.5$ and $I_0 \sim 0$), with one infected individual
  at $t=0$. Here the refractory individuals have phases $\tau_{i,j} =
  \tau_I + 1$ (namely, they are at the start of the refractory stage
  of the disease cycle). Again, the long bar shows the relative
  lengths of the susceptible (S), infected (I) and refractory (R)
  stages in the disease cycle, where $\tau_I=4$, $\tau_R=9$ and the
  total disease cycle $\tau_0$ is $13$ (see text). The red box shows
  the fraction of S, I and R individuals in the population at that
  instant of time. Interestingly, the spatially random population
  evolves into a more regular pattern after a short transient time. }
\label{spread_S=R}
\end{figure}

To begin with, in Fig. \ref{spread_S=R}, we illustrate the effect of a
single infected individual on an initial population with equal numbers
of susceptible and refractory individuals, namely $S_0 = R_0$.  It is
evident from these representative results that in a well mixed
population, consisting of a random collection of both susceptible and
refractory individuals, introduction of a single infected individual
can lead to {\em persistent infection in the population.}

  This can be rationalized as follows: the mixed presence of
  susceptible and refractory individuals, implies that the disease
  cycles of the individuals in the population are {\em not
    synchronized}. So there are always some individuals in the
  infective stage of the disease cycle in the population, and these
  act as seeds for continued infection propagation, leading to
  persistent infection. {\em Counter-intuitively then, the presence of
    individuals who are (temporarily) immune to the disease amongst
    susceptible ones leads to sustained infection, while in a
    completely susceptible population the infection dies out.}

%Note that these results consider the refractory individuals to have
%same phase in the disease cycle i.e. $\tau_R = \tau_I + 1$
%initially. In subsequent sections we will extrend the analysis to the
%case where the refractory individuals ($\tau_R$) have a randomly
%chosen phase from the range $[(\tau_I + 1), \tau_0]$ at initial
%time. We will show that even for the case of heterogeneous populations
%with random disease phase distributions, inclusion of an infected
%individual leads to persistent infection.

%\subsection{Time evolution of the infected sub-population}

%\begin{equation}
%I_t = \frac{\mbox{Total number of infected individuals at time $t$}}{\mbox{Total number of individuals in the population}}
%\end{equation}

%\begin{figure}[htb]
%	\includegraphics[width=0.45\textwidth]{fig5_time-seed.png}
%%{Fig5-time.png}
%        \caption{Time evolution of $I_t$, $S_t$, $R_t$, in a
%          heterogeneous population comprising initially of a random
%          mixture of individuals, with $S_0 = R_0$ and one infected
%          individual.}
%\label{time}
%\end{figure}

% The sustained emergence of infection from a center of infection and
% subsequent removal at the boundaries, explains the steady values
% $I_t$, $R_t$ and $S_t$.

\begin{figure}[htb]
\includegraphics[width=1\textwidth]{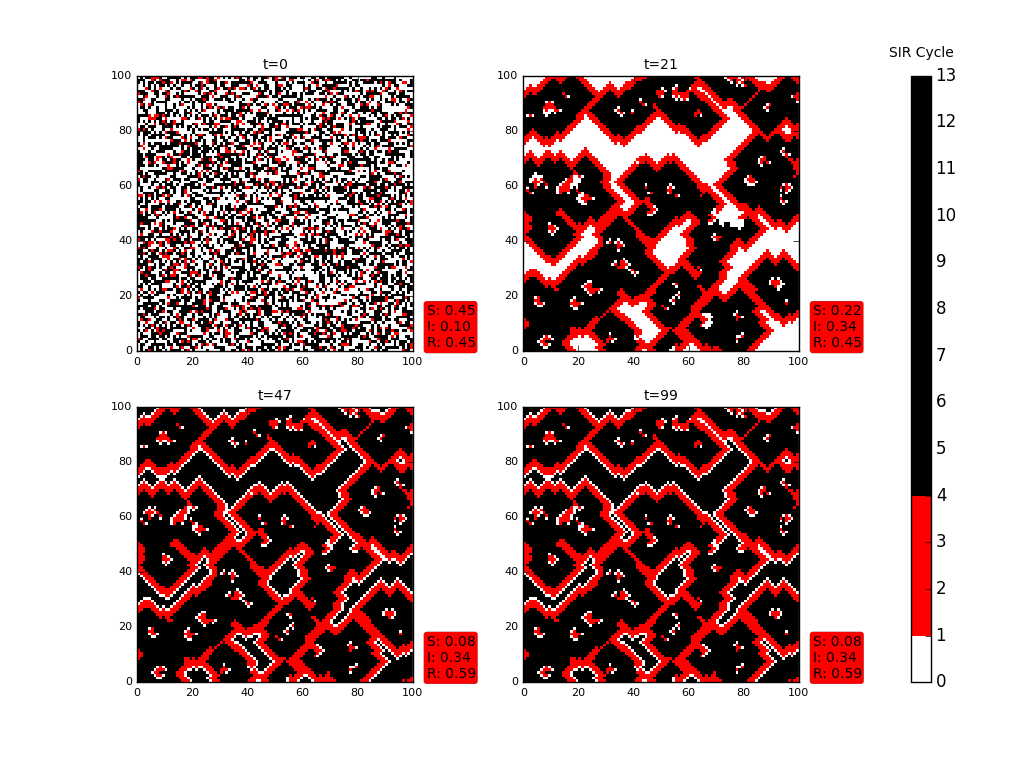}
%{Fig6-I0_01.png}
\caption{Snapshots of the infection spreading pattern in a
  heterogeneous population comprising initially of a random mixture of
  individuals, with $S_0 = R_0$ and $I_0=0.1$. Here the refractory
  individuals have phases $\tau_{i,j} = \tau_I + 1$ (namely, they are
  at the start of the refractory stage of the disease cycle). Again,
  the long bar shows the relative lengths of the susceptible (S),
  infected (I) and refractory (R) stages in the disease cycle, where
  $\tau_I=4$, $\tau_R=9$ and the total disease cycle $\tau_0$ is $13$
  (see text). The red box shows the fraction of S, I and R individuals
  in the population at that instant of time.}
\label{time1}
\end{figure}

%[Addition]
Next we focus on the time evolution of an initial population consisting
of a random mixture of $S$, $I$ and $R$ states. In particular we
investigate the nature of the persistent infection in the population
under varying initial fractions of infected individuals $I_0$.  A
typical random initial condition is shown in Fig. \ref{time1}, with
the initial fraction of infected sites $I_0$ being one-tenth and the
initial fraction of susceptibile and refractory individuals being
equal (i.e. $S_0 = R_0$).  Here too we find that infection is
sustained.

\begin{figure}[htb]
	\includegraphics[width=0.6\textwidth]{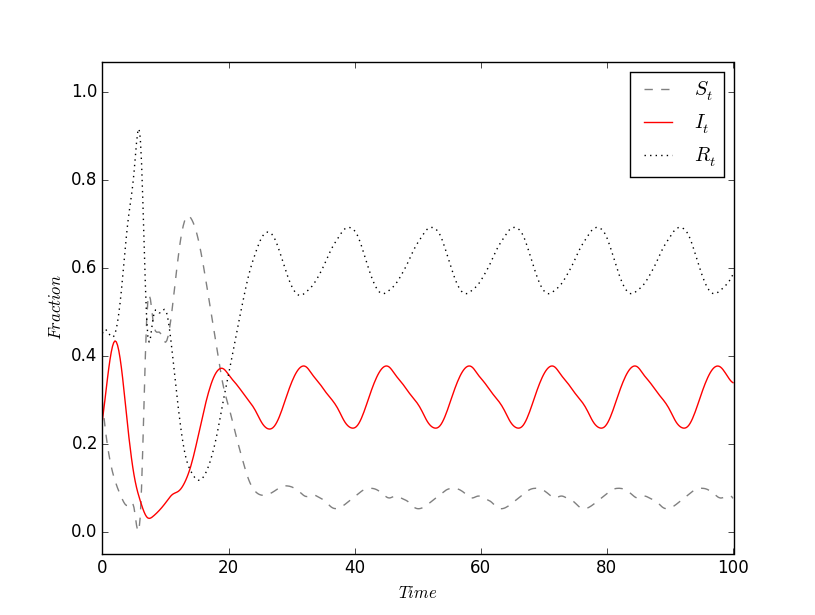}
        \caption{Time evolution of $I_t$, $S_t$, $R_t$, in a
          heterogeneous population comprising initially of a random
          mixture of individuals, with $S_0 = R_0$ and $I_0=0.1$.}
\label{time2}
\end{figure}

%The qualitative nature of the spread of infection and the progression
%of disease in the individuals with the passage of time, as discussed
%above, gives clear indication of the existence of persistent infection
%in a heterogeneous population. Now we present a quantitative analysis
%of the system to understand more precisely the nature of persistent
%infections.

Further, interestingly, it is clear that there is an {\em approximate
  recurrence of the complex patterns of infected individuals in the
  population}. Fig. \ref{time2} shows the time evolution of the
fraction of infected, refractory and susceptible individuals in the
population, namely $I_t$, $R_t$ and $S_t$, in the case displayed in
Fig. \ref{time1}. It can be clearly seen that after transience, $I_t$,
$R_t$ and $S_t$ exhibit steady oscillatory dynamics, with period of
oscillation close to the disease cycle length $\tau_0$. This is
consistent with the observed recurrence of the spatio-temporal patterns
when persistent infection emerges.

\begin{figure}[ht]
	\includegraphics[width=0.6\textwidth]{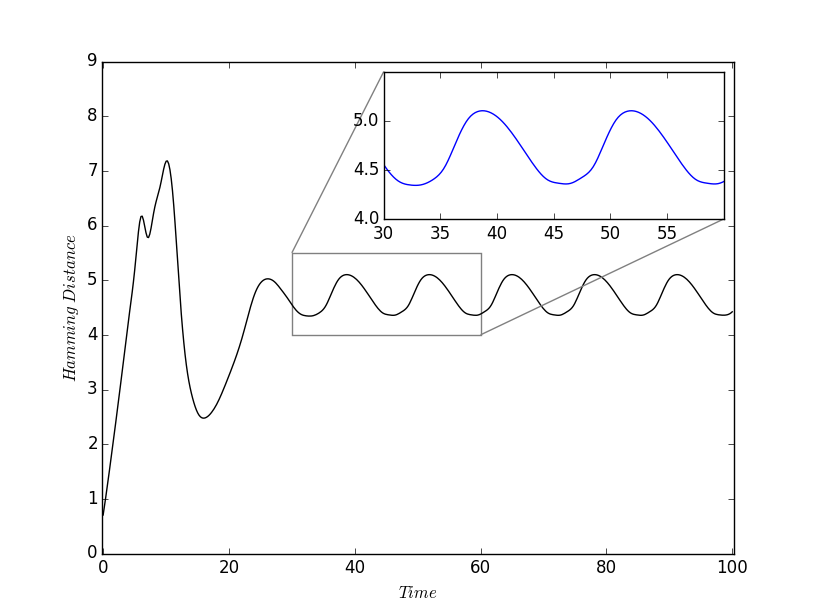}
        \caption{Hamming distance given by Eqn.~\ref{ham} as a
          function of time, in a heterogeneous population comprising
          initially of a random mixture of individuals, with $S_0 =
          R_0$ and $I_0=0.1$.}
\label{hamming}
\end{figure}

A quantitative measure of the recurrence of patterns can also be
obtained by calculating the difference of the state of the population
from the initial state, as reflected by the Hamming distance:
%, namely the lattice sum of the absolute value of the pair-wise difference of the states at the sites at time $t$ and initial time $0$.
\begin{equation}
H = \frac{1}{N} \sum_{i,j}\lvert \tau_{i,j}(t) - \tau_{i,j} (0) \rvert
\label{ham}
\end{equation}
where the sum is over all $N$ sites in the lattice. The time dependence
of the Hamming distance given above is shown in Fig. \ref{hamming},
and it clearly shows steady oscillations. This indicates that the
fraction of the susceptible, infected and refractory individuals in
the population, and more remarkably their {\em locations}, repeat
almost periodically over time. It should be noted that the frequency
of oscillation again approximately corresponds to the disease cycle
length.

%Although the oscillations are seen due to the difference in the
%physical dimensions of the popualtion lattice and infection spread
%pattern.

Another pertinent observation here is the dependence of this dynamics
on disease cycle. As the length of the infectious stage
(i.e. $\tau_I$) increases, keeping the total disease cycle length
invariant, the fraction of infected individuals $I_t$ increases. The
average $I_t$ is proportional to the fraction of the disease cycle
occupied by the infectious stage, i.e the ratio $\tau_I/\tau_0$. So
the size of the infected population strongly depends on the nature of
disease as reflected in the length of the infectious stage of the
disease.

\section{Influence of the initial composition of the population on the  persistence of infection}

We now attempt to gauge the statistically significant trends in $I_t$,
by averaging the fraction of infected individuals at asymptotic time
$t$,
%(where $t$ is significantly later than the transient stage),
arising from a wide range of initial configurations at time $t=0$. We
denote this by $\langle I_t \rangle$. In terms of this quantity,
persistent infection is indicated by a non-zero value. However,
after sufficient transient timesteps, if $\langle I_t \rangle$ is zero,
it indicates that the infection has died out. So $\langle I_t \rangle$
can serve as an order parameter for the transition to sustained
infection in a population.

\subsection{Dependence of persistence of infection on the initial fraction of susceptibles}

For fixed $\tau_I$ and $\tau_0$ we have calculated $\langle I_t
\rangle$, for different initial fractions of susceptible individuals
$S_0$. We explore the full possible range of $S_0 \in [0,1]$, where
$S_0=0$ signifies a population comprised entirely of refractory
individuals who are immune to infection initially, and $S_0=1$ implies
an initial population comprised entirely of individuals susceptible to
infection. While the phase of the susceptible (S) sub-population is
$\tau_{i,j} = 0$ of course, the refractory individuals (R) can be
present in different stages in the refractory period with $\tau_I <
\tau_{i,j} < \tau_0$.  We explore two different scenarios of the
initial state of the refractory individuals in the population.

%\item All refractory individuals are in the same phase.

%\item The refractory individuals have random phases, i.e. the
%  individuals may be at different stages of the refractory period.

%\end{itemize}

\begin{figure}[ht]
  \includegraphics[width=0.65\textwidth]{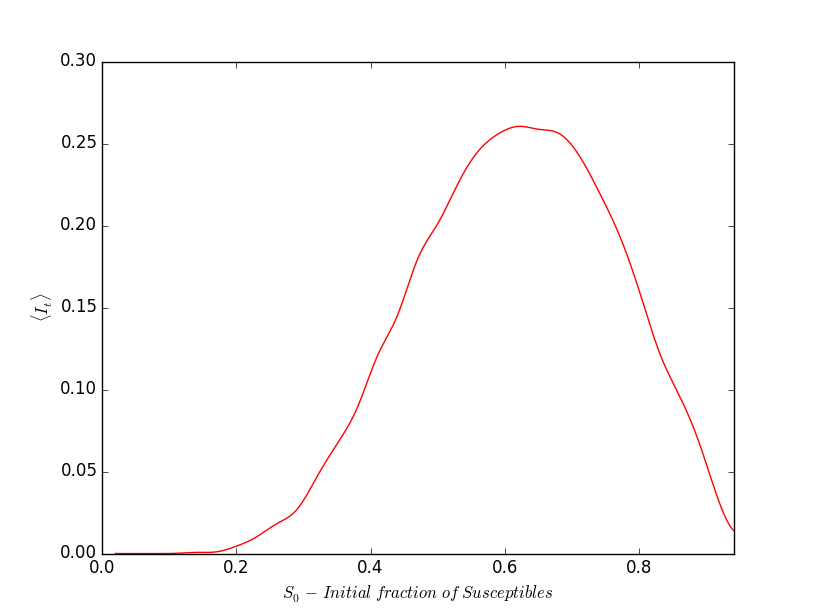}
%{S_0-v-I_t-rand.png}
%{variation-1.JPG}
  \caption{Variation of $\langle I_t \rangle$ (after transience) with
    respect to the fraction of susceptible individuals in the initial
    population $S_0$, arising from the presence of a single infected
    individual at time $t=0$. Here the refractory individuals have the
    same phase, the disease cycle has $\tau_I = 4$; $\tau_0 = 13$, and
    $I_t$ is averaged over $10^3$ realizations of the initial
    population on the lattice. The specific case of a $100 \times 100$
    lattice is displayed. However note that different lattice sizes
    yield the same result.}
\label{variation1}
 \end{figure}

 %We present the specific case in Fig. \ref{variation1}, of all
 %$\tau_{i,j} = \tau_I + 1$ for the uniform phase case, and
 %$\tau_{i,j}$ randomly chosen from the interval [$\tau_I + 1, \tau_0$]
 %to illustrate the scenario of random phase distributions.

 First we present the case where all the refractory individuals are at
 the start of the refractory stage of the disease cycle, i.e. all
 $\tau_{i,j} = \tau_I + 1$. So there is uniformity in the stage of
 disease progression in the refractory sub-population, though the
 individuals are randomly distributed spatially. We focus on the
 asymptotic state of infection in such a population, arising from a
 single infected individual at the outset. The results obtained from a
 large sample of initial states is shown in Fig. \ref{variation1}, and
 it is evident from there that $\langle I_t \rangle$ is {\em very low
   for both high and low $S_0$}, peaking around $S_0 \sim
 0.65-0.75$. Namely, homogeneous initial populations where all
 individuals are immune ($S_0 = 0$), or all are susceptible to disease
 ($S_0 = 1$), do not yield persistent infection. Rather, mixed
 populations lead to most sustained infection, with persistently high
 numbers of infected individuals.

 We can rationalize our observations as follows:
% here all the individuals in the refractory phase initially have same
% phase $\tau_R = \tau_I + 1$.
 If an infected individual is completely surrounded by refractory
 individuals with $\tau = \tau_I+1$, it will complete the infectious
 stage without transferring the infection at all, as $\tau_I <
 \tau_R$. So the infection can spread only if the infected seed is
 contiguous to at least one susceptible individual. Now the
 probability of contact with a susceptible individual in the initial
 stages of infection spreading depends on the initial fraction of
 susceptibles $S_0$. This suggests that when $S_0$ is low, the chance
 of the infected individual being in contact with a susceptible one is
 low. As a result, as $S_0$ tends to zero, on an average, the
 infection eventually gets removed from the population, with the seed
 of infection crossing over to the refractory phase without infecting
 any other individual.

 When there are more susceptible individuals in the initial
 population, there is a higher chance that the infected seed will
 encounter a susceptible neighbour. So as expected, increasing $S_0$
 leads to a larger infected set on an average. However the surprising
 trend is the {\em decrease} in the infected set as the initial
 susceptible sub-population becomes too high, with the number of
 infected individuals tending to zero as the entire population becomes
 susceptible. This feels counter-intuitive, but can be understood as
 follows: Consider the limiting case where initially almost all the
 individuals are susceptible to the infection. Now the infection will
 spread immediately in isotropic waves, but will eventually stop at
 the boundaries. In analogy to the spread of forest fire, the boundary
 of refractory individuals is like scorched earth preventing spread
 across them. Now after the wave of infection passes, the individuals
 are in the refractory stage, leading eventually to the entire set
 being synchronized in the susceptible regime. There is no infected
 individual left then to act as a seed for a further wave of infection
 spreading. So the infection does not persist. The susceptible stage
 is like an ``absorbing state'', and in the absence of ``infectious
 perturbation'' the system remains fixed in that state.

 In order to prevent the above scenario, one needs enough refractory
 individuals in the population. When $R_0$ is below $1/4$ (i.e. $S_0 >
 3/4$), typically the infected seed may not have a refractory
 individual among its four neighbours. So one expects that the
 persisting infection will have lower probability of occurrence as
 $S_0$ increases beyond $3/4$. This is in accordance with the trends
 observed in the simulations.

 We then see that for the {\em infection to persist} in a population,
 a {\em well mixed heterogeneous population is required}, with
 reasonable number of both susceptible and refractory
 individuals. {\em Randomly mixed populations prevent synchronization
   of the disease}, and this is the key to always having some source
 of infection left in the population.

\subsection{Dependence of persistence of infection on the initial fraction of infecteds}

\begin{figure}[ht]
  \includegraphics[width=0.75\textwidth]{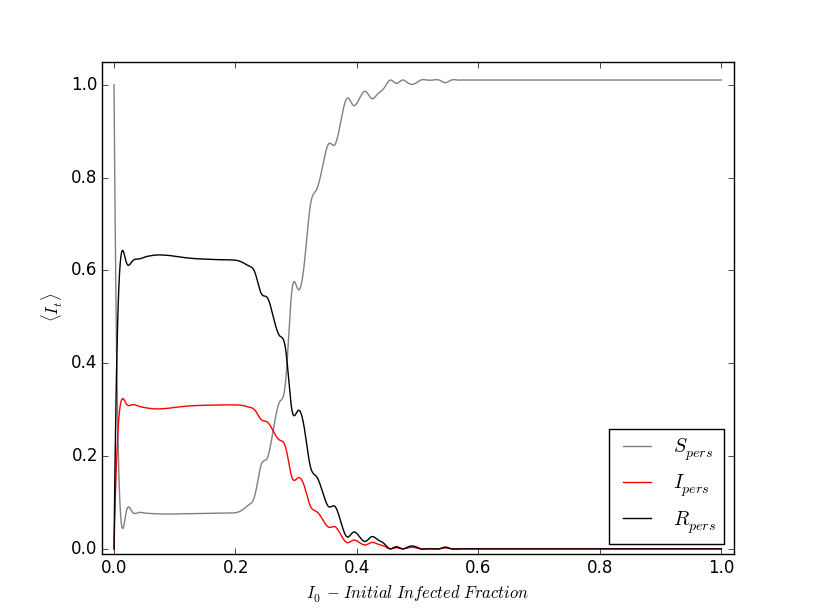}
%{Fig10-perst_s=r.png}
  \caption{Variation of $\langle I_t \rangle$ (after transience) with
    respect to the initial fraction of infected individuals $I_0$ in
    the population, and $S_0=R_0$. The refractory sub-population
    consists of individuals with phase equal to $\tau_I+1$. Here the
    disease cycle has $\tau_I = 4$; $\tau_0 = 13$, and $I_t$ is
    averaged over $10^3$ initial realizations. The specific case of a
    $100 \times 100$ lattice is displayed. However note that different
    lattices sizes yield the same result.}
\label{I0}
 \end{figure}

We now vary the initial fraction of infected individuals $I_0$ in the
population, over the entire range $[0,1]$. For the
remaining population, the initial fraction of susceptible and
refractory individuals is set at different ratios.
% such as $S_0:R_0 =1:1$ or $2:1$ or $1:2$.
%For each fraction $I_0$, 100 specific instances of random initial conditions are generated, each having a different arrangement of the $S$, $I$ and $R$ states, but with identical composition corresponding to the fraction $I_0$. In all $10^4$ random initial conditions are generated.\\
%Each of the initial conditions of a particular initial fraction are allowed to evolve in time, long enough for the transients to pass, and then observed for a fixed time period. The fraction of infected individuals that are present at each time step are recorded.\\
%The number of infected individuals is then averaged over the time period of observation, for a single system. This gives the time average of persistence of the infection for the single instance of the ensemble corresponding to a single $I_0$ fraction (note that each instance of the ensemble has identical $S_0$, $I_0$ and $R_0$). %The time averaged persistence values obtained in the previous step for all the instances of the ensemble are then averaged, to obtain a single value $\langle I_t \rangle$ for each fraction in the set $[0,1]$. The results are presented below.\\
We consider an ensemble of initial conditions, with specific $I_0$,
$S_0$ and $R_0$ and find the time averaged $I_t$, after long
transience for each realization. The ensemble average of this quantity
is displayed in Fig. \ref{I0}. Notably, we find that there is a
definite {\em window of persistence} over the range of $I_0$, where
the infection never dies down and the fraction of infected individuals
in the population is reasonably high on an average.

 In the state where infection is persistent, the individuals are
 unsynchronized and spread over the different stages of the disease
 cycle. So on an average the fraction of infected individuals is $\sim
 \tau_I/\tau_0$, namely the fraction of the total disease cycle
 occupied by the infected stage. For instance, in the example shown in
 Fig. \ref{I0} with $\tau_I=4$ and $\tau_0=13$, at the plateau of
 persistence, the infected fraction is approximately one-third of the
 population.

 The transition to persistent infection is sharp and occurs at $I_0
 \rightarrow 0$. This implies that {\em the infection can spread and
   persist even when there is only a single infected individual in the
   initial population}. This is consistent with the results we
 presented earlier (cf. Fig. \ref{variation1}) on infection spreading
 from a single infected individual.

 Interestingly, the infection ceases to persist for higher values of
 $I_0$, and the fall in persistence is rapid. That is, if the initial
 population has too many infected individuals, infection will not
 persist. This can be rationalized by noting that one needs a mix of
 susceptibles and refractory individuals in the population for persistent
 infection. For instance, considering the limiting case of all
 infected individuals in the initial state, it is clear that all
 individuals will go through the disease cycle in synchrony. So all
 individuals will become susceptible again together, but there will be
 no infective seed left in the population to perpetuate the infection.

 \section{Effect of varying degrees of non-uniformity in the
   refractory sub-population on the persistence of infection}

 Now we will explore the effect of non-uniformity within the
 refractory sub-population on the emergence of persistent
 infections. Namely, we will consider the refractory individuals in the
 initial population to be in different stages of disease
 progression. We will consider two distinct ways of interpolating
 between the completely heterogeneous and completely uniform limiting
 cases, in order to gauge the effect of heterogeneity on sustaining
 infection.

%\subsubsection{Variation of the fraction of initial refractory sub-population having random phases $(\eta):$}

 First we consider the initial refractory sub-population to be an
 admixture of subsets of individuals with uniform phase and with
 randomly distributed phases. Specifically, we explore initial
 refractory sub-populations comprised of some fraction $f_{rand}$ with
 phases randomly distributed over the range $\tau_I + 1$ to $\tau_0$,
 and the rest $1-f_{rand}$ with fixed phase $\tau_R = \tau_I + 1$.  We
 examine the spread and persistence of infection in such a scenario,
 under variation of the initial composition of the population.

\begin{figure}[ht]
  \includegraphics[width=0.75\textwidth]{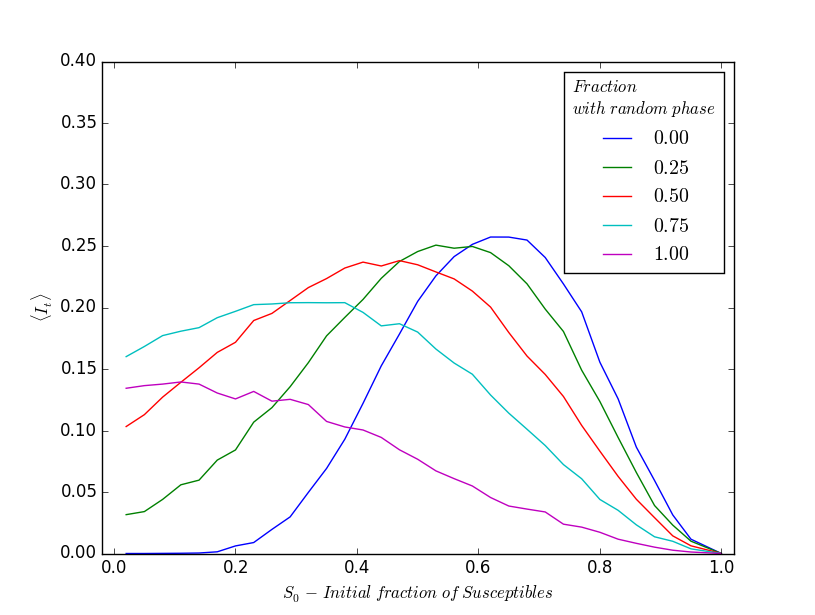}
%{fig11_S-0-trend-n.png}
%{transition-population.jpg}
  \caption{ Variation of $\langle I_t \rangle$ (after transience) with
    respect to initial fraction of susceptible individuals $S_0$, for
    different fractions $f_{rand}$ of the initial refractory
    sub-population having randomly distributed phases (see key). Here
    the disease cycle has $\tau_I = 4$; $\tau_0 = 13$, and $I_t$ is
    averaged over $10^3$ initial realizations and lattice size is $100
    \times 100$.}
\label{transition-p}
 \end{figure}

 Fig~\ref{transition-p} exhibits the persistence of infection, with
 respect to varying $S_0$, arising in a population that had a single
 infected individual initially. Different fractions of the initial
 refractory sub-population with randomized phases were explored,
 ranging from $f_{rand}=0$ (i.e. completely uniform), to $f_{rand} =1$
 (i.e. completely heterogeneous). The trends clearly indicate a
 continuous cross-over from the condition where all refractory
 individuals are in the same phase, to the scenario where all are in
 random phases.

%\begin{figure}[ht]
%  \includegraphics[width=0.65\textwidth]{S_0-v-I_t.png}
%{variation-1.JPG}
%  \caption{Variation of $\langle I_t \rangle$ (after transience) with
%    respect to the initial fraction of susceptible individuals in the
%    population $S_0$. Here the refractory sub-population consists of
%    individuals with randomly distributed phases, the disease
%    cycle has $\tau_I = 4$; $\tau_0 = 13$, and $I_t$ is averaged over
%    $10^3$ initial realizations. The specific case of a $40 \times 40$
%    lattice is displayed. However note that different lattices sizes
%    yield the same result.}
%\label{variation1}
% \end{figure}

\begin{figure}[ht]
  \includegraphics[width=0.75\textwidth]{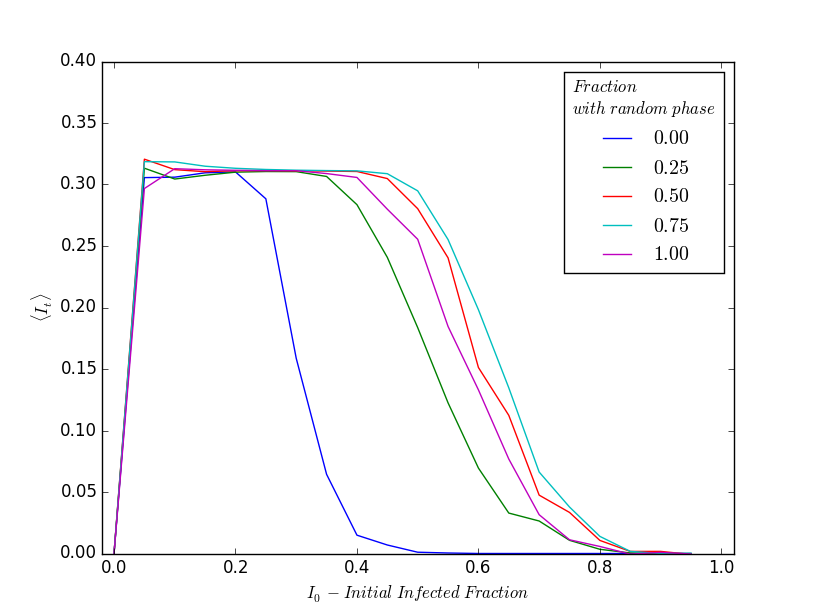}
%{Fig11-perst_trend_r_n.png}
  \caption{Variation of $\langle I_t \rangle$ (after transience) with
    respect to the initial fraction of infected individuals $I_0$ in
    the population, and $S_0=R_0$. The initial refractory sub-population
    consists of different fractions $f_{rand}$ with randomly
    distributed phases (see key). Here the disease cycle has $\tau_I
    = 4$; $\tau_0 = 13$, and $I_t$ is averaged over $10^3$ initial
    realizations. While the specific case of a $100 \times 100$ lattice
    is displayed, different lattices sizes yield the same result.}
\label{I0_n}
 \end{figure}

 Further, we explore the effect of varying the initial fraction of
 infected individuals $I_0$, over the range $[0,1]$. Fig. \ref{I0_n}
 exhibits the change in the window of persistence with respect to
 $f_{rand}$. It is evident that increasing $f_{rand}$, namely
 increasing the initial number of refractory individuals with {\em
   de-synchronized phases}, leads to a definite increase in the window
 of persistence. This implies that {\em for populations with a more
   heterogeneous refractory sub-population, the disease persists over a
   larger range of infected fractions $I_0$ of the initial
   population}.

 Note however, that there is also an apparent reduction in the window
 of persistence at very high $f_{rand}$.  This can be rationalized by
 noting that when the entire initial refractory sub-population $R_0$
 has uniformly distributed phases, there are a significant number of
 individuals who are closer to the end of their disease cycle (for
 instance, stage $12$ or $13$). These individuals become susceptible
 within a few time steps, and therefore bring the population closer to
 an overall state of homogeneity again, as all susceptibles are in the
 same phase (stage 0) and remain in that phase unless infected. We
 have observed qualitatively and quantitatively earlier, that a more
 homogeneous population leads to a reduced window of
 persistence. Hence, {\em presence of a significant number of
   individuals closer to the end of their disease cycle acts as a
   homogenizing factor for the population and is detrimental to
   persistence}.

%\subsubsection{Variation of the range of the random phases of the refractory sub-population $(\chi):$}

 Lastly, we study the effect of varying ranges of spread in the
 initial phases of the refractory individuals. Specifically we
 consider that the phase of the refractory individuals in the initial
 population to be randomly distributed over different ranges
 $R_{rand}$. In particular we examine the persistence of infection for
 $R_{rand}$ ranging from $[ \tau_I +1, \tau_I +1 ]$, (where all
 refractory individuals have the same phase) to [$\tau_I + 1$, $\tau_I
 + \tau_R$] (where heterogeneity is large as the phases of the
 refractory individuals are distributed over the entire refractory
 range).

 Fig. \ref{transition-r-1} exhibits representative results of $\langle
 I_t \rangle$ as a function of the initial fraction of susceptibles
 $S_0$, for the case where there is a single infected individual in
 the population at the outset. It can be clearly seen that a smooth
 cross-over takes place from the extremal case of all refractory
 individuals in the same phase, to the limit where the stages of the
 refractory individuals are spread randomly over the entire refractory
 period. The key observation here is that as the spread in phases
 increases, the range of persistent infection becomes larger. Namely,
 when there is a large initial spread in the stages of disease among
 the individuals, at subsequent times there are always some
 individuals who can ``pick up the baton of infection'',
% from those that were susceptible in the early stage of disease spreading,
leading to persistent infection.

\begin{figure}[ht]
\includegraphics[width=0.75\textwidth]{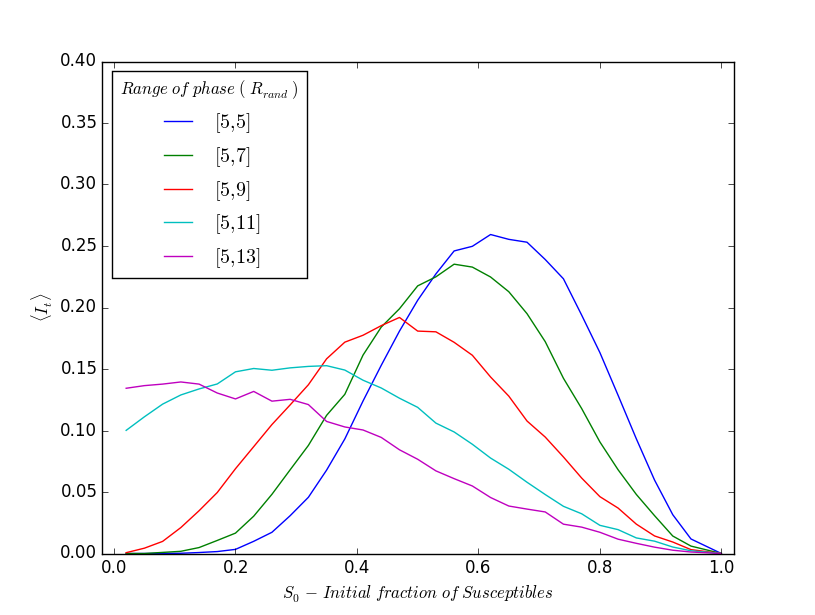}
%{fig13_S-0-trend-x.png}
%{transition-range-1.jpg}
%\includegraphics[width=0.65\textwidth,height=0.45\textwidth]{transition-range-2.JPG}
\caption{ Variation of $\langle I_t \rangle$ (after transience) with
  respect to initial fraction of susceptible individuals $S_0$, for
  the refractory individuals having phases $\tau$ randomly distributed over
  different ranges $R_{rand}$ in the refractory stage : [5,5];
  [5,7]; [5,9]; [5,11]; [5,13].
% (bottom) [5,13]; [7,13]; [9,13].
  Here $I_t$ is averaged over $10^3$ realizations, lattice size is $100
  \times 100$, and the disease cycle parameters $\tau_I = 4$, $\tau_0 =
  13$.}
\label{transition-r-1}
\end{figure}

\begin{figure}[ht]
\includegraphics[width=0.65\textwidth,height=0.45\textwidth]{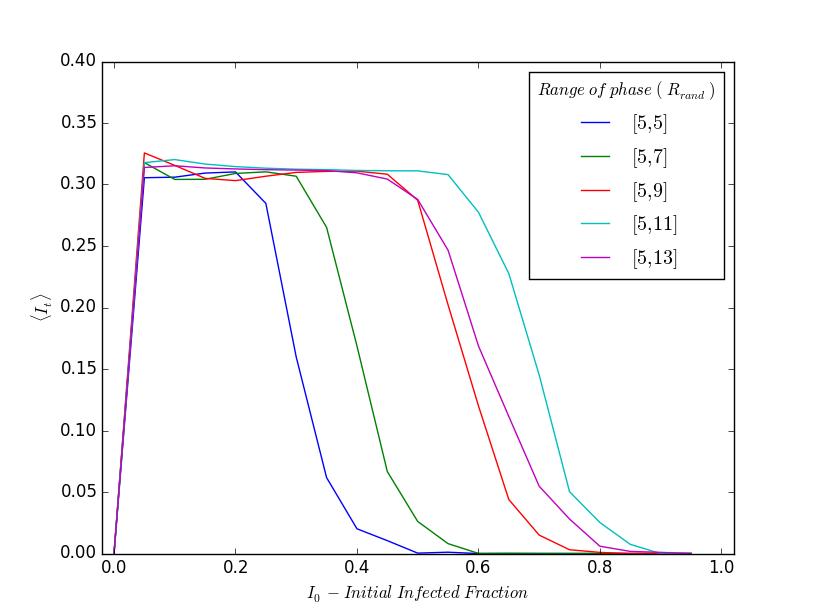}
%{transition-range-1.jpg}
%\includegraphics[width=0.65\textwidth,height=0.45\textwidth]{transition-range-2.JPG}
\caption{ Variation of $\langle I_t \rangle$ (after transience) with
  respect to initial fraction of infected individuals $I_0$, for
  the refractory individuals having phases $\tau$ randomly distributed over
  different ranges $R_{rand}$ in the refractory stage : [5,5];
  [5,7]; [5,9]; [5,11]; [5,13].
% (bottom) [5,13]; [7,13]; [9,13].
  Here $I_t$ is averaged over $10^3$ realizations, lattice size is $100
  \times 100$, and the disease cycle parameters $\tau_I = 4$, $\tau_0 =
  13$.}
\label{transition-r-1}
\end{figure}

% Lastly, we consider refractory phases randomly distributed around a
% mean, for instance around the middle of the refractory range.

% The initial fraction of susceptible and refractory individuals that
% yield the most pronounced persistent infection (namely, the value of
% $S_0$ for which $I_t$ is maximum) shows a smooth increase with
% decreasing heterogeneity in the population. So it is clear that
% heterogeneity in the individual disease cycle phases plays a key
% role in the persistent infection as well.

% However, if the refractory individuals initially have random phases,
% then even for very low initial $S_0$ the number of infected
% individuals after some time is large. Further this number decreases
% monotonically in the range $S_0 \in [0,1]$. These results are
% quantitatively the same for different lattice sizes.
So we see that in the completely heterogeneous case, low susceptible
and high refractory initial subpopulations favour persistent
infection. But in a completely uniform population, a higher fraction
of susceptibles leads to persistent infection. This has the following
important implication: when refractory individuals are not
synchronized at the same phase of disease progression, even if there
are few susceptible individuals in the population initially, the
infection grows substantially and the average size of the infected
sub-population is large.  So we have demonstrated that even when the
entire population is susceptible to infection, the infection
eventually dies out, while even a few susceptibles among an
heterogeneous refractory population gives rise to a large persistent
infected sub-population.

%% Varying $I_0$ over $[0,1]$ with fixed proportions of $S_0$ and $R_0$:

We can rationalize this counter-intuitive trend that persistent
infection is more likely when the number of susceptible individuals in
the initial population is low, as follows:
% First, these trends are consistent with the limiting case of
% completely susceptible populations where persistent infection does
% not emerge.
When $S_0$ is low, there are many refractory individuals in the
population surrounding the infected individual. These individuals are
in various stages in the refractory period, and some become
susceptible again while the seed is still infectious. If $S_0
\rightarrow 0$ and the refractory individuals are uniformly
distributed over the refractory range $\tau_R$, the probability of the
seed encountering a susceptible individual while still infectious is
proportional to $\tau_I/\tau_R$. Since at least one neighbour in
contact with the seed needs to be susceptible, this probability should
be greater than $\frac{1}{4}$ for the infection to spread, on an
average. So when the infective stage $\tau_I$ is sufficiently long (as
in our example of $\tau_I =4$, in a disease cycle of length $13$),
extremely low initial $S_0$ can also lead to persistent infection.
% At lower range of $S_0$, considerably higher $I_t$ values represents
% that for a population spread with larger proportion of refractory
% individuals (in randomly different phases) has a higher chance of
% persistent infection due to the inclusion of infected individual in
% the population.
% Consider a situation of population spread with low $S_0$ and
% persistent infection is seen. The infected individual is introduced
% to the population initially, as the infection period is as long as
% $\tau_I$ (taken to be 4 here) so the individual will remain infected
% for this duration. The susceptible neighbors of this infected
% individual then gets infected and infection propogates. Now if the
% neighbor is refractory with its phase towards the end of the
% refractory period, then within the time $\tau_I$ it becomes
% susceptible and consequently gets infected. So the infection
% propogates in this case as well, and the infection is sustained as
% there are also refractory individuals with phase closer to the
% beginning of the refractory period.

\section{Conclusion}

In summary, we have explored the emergence of persistent infection in a
patch of population, where the disease progression of the individuals
was given by the SIRS model and an individual became infected on
contact with another infected individual. We investigated the
infection spreading qualitatively and quantitatively, under varying
degrees of heterogeneity in the initial population.

Specifically, we considered two scenarios extrapolating between the
completely homogeneous and completely heterogeneous limit. One
considers varying fractions of heterogeneous sub-populations and
another examines varying ranges in the spread of the stages of disease
progression. Our central result is the following: we find that an
infectious seed does not give rise to persistent infection in a
homogeneous population consisting of individuals at the same stage of
disease progression. Rather, when the population is heterogeneous, and
consists of randomly distributed individuals at various stages of the
disease, infection becomes persistent in the population patch. The key
to persistent infection is then the random admixture of refractory and
susceptible individuals, leading to de-synchronization of the phases
in the disease cycle of the individuals.  So we have demonstrated that
when the entire population is susceptible to infection, the infection
eventually dies out, while even a few susceptibles among an
heterogeneous refractory population gives rise to a large persistent
infected sub-population.

\bigskip
\bigskip

{\bf Author contributions}\\

SS conceived the problem, and VA and PM performed all the numerical
simulations. SS, VA and PM discussed the results and wrote the
manuscript together.\\

\bigskip

The authors declare no competing financial interests.\\

\end{document}